\documentclass[prd,twocolumn]{revtex4}

\usepackage{epsfig}
\usepackage{hyperref}

\newcommand{\bvect}[1]{\ensuremath{\mathbf{#1}}}
\newcommand{\vect}[2]{\ensuremath{#1 _{#2}}}
\newcommand{\pder}[2]{\ensuremath{\partial_{#2} #1}}
\newcommand{\pdert}[3]{\ensuremath{\partial_{#3} \partial_{#2} #1}}
\newcommand{\fder}[2]{\ensuremath{\frac{d #1}{d #2}}}

\newcommand{\fpder}[2]{\ensuremath{\frac{\partial #1}{\partial #2}}}
\newcommand{\vder}[2]{\ensuremath{\frac{\delta #1}{\delta #2}}}
\newcommand{\diverge}[2]{\ensuremath{\pder{\vect{#1}{#2}}{#2}}}
\newcommand{\laplace}[2]{\ensuremath{\partial_{#2}^{2} #1}}

\newcommand{\sci}[2]{\ensuremath{#1 \times 10^{#2}}}

\newcommand{\norm}[1]{\ensuremath{\left\| #1 \right\|}}
\newcommand{\normtwo}[1]{\ensuremath{\norm{#1}_{2}}}

\newcommand{\td}[2]{\ensuremath{#1_{#2}}}

\newcommand{\intsq}[1]{\ensuremath{\overline{#1}}}

\newcommand{\eqnref}[1]{(\ref{eqn:#1})}
\newcommand{\secref}[1]{Section~\ref{sec:#1}}
\newcommand{\figref}[1]{Figure~\ref{fig:#1}}
\newcommand{\tabref}[1]{Table~\ref{tab:#1}}

\begin{document}

\begin{abstract}
There is an abundance of empirical evidence in the numerical relativity 
literature that the form in which the
Einstein evolution equations are written plays a significant role in the
lifetime of numerical simulations.  This paper attempts to present 
a consistent framework for modifying
any system of evolution equations by adding terms that
push the evolution toward the constraint hypersurface.  
The method is, in principle, applicable to any system
of partial differential equations which can be divided into evolution equations
and constraints, although it is only demonstrated here through an 
application to the Maxwell equations.
\end{abstract}

\title{Toward Making the Constraint Hypersurface an Attractor in Free
Evolution} 
\author{David R. Fiske}
\email{drfiske@physics.umd.edu}
\affiliation{Department of Physics, University of Maryland,
College Park, MD 20742-4111}

\maketitle

\begin{section}{Introduction}
The data analysis needs of the LIGO and LISA experiments have driven
an interest in finding numerical solutions to the Einstein equations in
astrophysically relevant situations.  Wave forms from such numerical 
simulations can, once generated, be used as templates for finding
gravitational wave signals in the experimental noise.

In order to find numerical solutions, however, one must make a 3+1 space-time
split of the Einstein equations following some variation of the ADM 
formalism \cite{adm}.
When using free evolution techniques, which is the most common practice in
numerical relativity today,
a plethora of authors have found that the exact form of the evolution
equations, in particular how the constraints are substituted into the
evolution equations, plays a critical role in how stable the
evolution will be 
(e.g. \cite{sn:bssn,bs:bssn,aei:bssn,kst:hyperbolic,lugshoe:mixed}).  
Although making such modifications to the evolution equations remains more 
of an art than a science, some authors have proposed well defined methods for 
making stability-friendly modifications.

Historically, Detweiler \cite{det:mod} 
seems to be one of the first authors to consider explicitly a method
for adding terms proportional to the constraints to the evolution equations
with the goal of improving simulation stability.
He looked at the evolution of the constraints, and
tried to add terms to the evolution equations of the fundamental variables
that would drive the constraints to zero as the evolution progressed.  This
approach is very similar to the approach presented here.  

Where the approach differs,
however, is significant.  Whereas Detweiler needed to add terms such that
a negative-definite operator appeared on the right hand side of
 the evolution equation for the
constraints, the method proposed here generates a general term of the
correct sign which is added to terms of unknown sign that come from
the underlying formalism.  Under special circumstances, Detweiler showed
that his method does in fact generate a negative-definite operator by making
use of a free parameter introduced in his correction terms.  The method
proposed here also uses a free parameter in arguing that the constraints
can be evolved toward zero, with the difference that the argument here
is more general but without a guarantee that the overall sign of the
the constraint evolution has the correct sign.

The $\lambda$-system approach, proposed by Brodbeck, Frittelli, H\"{u}bner,
and Reula, embeds 
the Einstein equations
into a larger system of symmetric hyperbolic equations.  The Einstein equations
are coupled to the larger system by adding terms to the Einstein evolution
equations which are zero on the constraint hypersurface, with the 
extra degrees of freedom defined such that they are forever zero if the
Einstein constraints and constraints on the new degrees of freedom 
are satisfied in the initial data.  A linear order analysis of this
system suggests that for ``small enough'' deviations from the constraint
hypersurface, the system should asymptote to a solution of the Einstein 
equations \cite{reula:lambda}.

Finally, Yoneda and Shinkai \cite{hisaaki:eigen} performed detailed 
eigenvalue calculations to determine the theoretical signs of the constraint 
evolution when linearizing around a Minkowski background spacetime.
While their program was intensive and thorough,
their results depend on the validity of a perturbation expansion around a
fixed spacetime.  In that limit, they evaluate the sign of the eigenvalues
of their expansion matrix, and use those signs to predict the goodness of
correction terms which are strictly linear in the constraints and
spatial derivatives of the constraints.  Although one of their
more promising terms was used to good effect in a numerical experiment
\cite{yo:octant}, their analysis remains strictly valid only in the
perturbation regime around Minkowski space, and, at present, they have only
examined terms linear in the constraints and derivatives of the constraints.

The method here, in contrast, attempts to make the 
constraint surface an attractor
for the evolution equations without appealing to perturbation theory
in any way and without extending the system of equations.  
The results will show it successful in reaching this goal
for the simple case of the Maxwell equations, but cannot confirm
with certainty that the results will generalize to non-linear general
relativity.  The primary problem is that the correction terms proposed 
here have the form
\[ \fder{}{t} \mathrm{constraint}^{2}
= (\mathrm{orig\ rhs}) - (\mathrm{correction}) \] 
with a correction term of definite sign subtracted from a term
of indefinite sign coming from the underlying theory.  The success of
the method in the case of the linear Maxwell equations makes a similar
result in the linearized Einstein equations likely, but does not 
guarantee success for full, non-linear general relativity.  This
will be discussed more completely in \secref{conclusions}.

In this paper, following earlier work by Knapp, Walker, and Baumgarte 
(KWB) \cite{baumgarte:em}, I will apply the method
to Maxwell's equations as a proof-of-concept test.  The work by KWB
showed that some qualitative features of two
popular ADM formulations of general relativity are manifest in the
simpler framework of electromagnetism, making this a natural place to
demonstrate the feasibility of the new methods presented here.
\end{section}

\begin{section}{The Theory}
\label{sec:theory}
I will introduce the idea here by first considering the ordinary differential
equation example of the simple harmonic oscillator, and then considering
the more interesting case of partial differential equations.

\begin{subsection}{ODE: Harmonic Oscillator}
The equations of motion for a harmonic oscillator 
(with $\omega = 1$) are 
\begin{eqnarray}
\fder{x}{t} & = &  v \\
\fder{v}{t} & = & -x
\end{eqnarray}
and the constraint \footnote{This is actually a conserved quantity.  For
this simple ODE case, there is no difference, but when we pass to PDEs, there
will be a difference between conserved quantities and constraints.} is the 
shifted energy function
\begin{equation}
E(t) = x^{2} + v^{2} - E_{0}
\end{equation}
with \(E_{0} = x^{2}(0) + v^{2}(0)\),
which should be zero-valued at all times.  In a numerical evolution, one
would like to evolve the fundamental quantities $x$ and $v$, and, after doing
so, calculate the constraint $E(t)$ from $x$ and $v$.  Since $E$
is derived, on the other hand, it makes sense to look at its time
evolution in terms of the time evolution of $x$ and $v$ using the chain rule.
Furthermore, $E(t)$ is not bounded on either side, so consider instead
$E^{2}(t)$, which should also always be zero.  Applying the chain
rule gives the elementary result that
\begin{equation}
\fder{E^{2}}{t} \equiv F(x,v) 
= \fpder{E^{2}}{x} \fder{x}{t} + \fpder{E^{2}}{v} \fder{v}{t}.
\label{eqn:harme}
\end{equation}
Now we cannot change the two partial derivative factors since they are fixed
by the form of the energy expression, but, as noted before, we are free
to change the equations of motion 
without changing the physics
by adding multiples of the constraints.
Let $K$ be a positive constant.  If we make the changes
\begin{eqnarray}
\fder{x}{t} & = & v - K \fpder{E^{2}}{x}
	\label{eqn:newharmx} \\
\fder{v}{t} & = & -x - K \fpder{E^{2}}{v}
	\label{eqn:newharmv}
\end{eqnarray}
to the equations of motion, then
\begin{equation}
\fder{E^{2}}{t} = F(x,v) 
	- K \left[ \left(\fpder{E^{2}}{x}\right)^{2}
	+ \left( \fpder{E^{2}}{v} \right)^{2} \right]
\end{equation}
shows how the constraint evolves under the new equations of motion.  For
large $K$, the term in brackets will dominate the evolution of the constraint,
constantly pushing it toward its minimum value of zero.

Numerical results, integrating forward in time with the Euler method,
are in agreement with this analytic result.
\end{subsection}

\begin{subsection}{PDEs}
Having seen the method in the simple case of ODEs, consider now
the the more interesting case of PDEs with one constraint equation.
Let $s$ be the state vector for a system.  Define
\begin{equation}
S_{m}(t,\bvect{x})  = \fpder{s_{m}}{t}
\end{equation}
to be the right hand side of the evolution equation in the
unmodified formalism.  
A general constraint $C$ will depend on $s$
and its spatial derivatives.  Furthermore, the constraint should be 
satisfied at every point in space.

These considerations motivate looking at the integrated, squared constraint
\( \intsq{C^{2}} = \int C^{2} d^{N}x\), and, 
instead of taking partial derivatives with
respect to the fields, we need to take variational derivatives of the 
integrated constraint when considering the analogies of 
Equations~(\ref{eqn:newharmx}--\ref{eqn:newharmv}) so that the dependence
of $C$ on the spatial derivatives of the fields is treated properly.

Following this prescription, the appropriate modification to the
equation of motion for the state vector is
\begin{equation}
\fpder{s_{m}}{t}  =  S_{m}(t,\bvect{x}) 
	- K_{mn}(t,\bvect{x}) 
	\vder{\intsq{C^{2}}}{s_{n}(t,\bvect{x})} 
\label{eqn:news}
\end{equation}
for some positive-definite matrix-valued function $K_{mn}$.
Under this change,
\begin{equation}
\fder{\intsq{C^{2}}}{t} = D[s]
- \int
	\left( \vder{\intsq{C^{2}}}{s_{m}} \right)
	 K_{mn}
	\left( \vder{\intsq{C^{2}}}{s_{n}} \right) d^{N}x
\label{eqn:gencorrection}
\end{equation}
gives the evolution of the constraint in the modified theory.
Here $D[s]$ gives the functional form of the
right hand side of the constraint's evolution equation in the
unmodified theory.  For the cases considered here, I chose
the $K_{mn}$ diagonal and constant.

For systems with $M$ constraint equations, the method is
easily modified by taking the grand constraint functional to
be
\begin{equation}
\intsq{C^{2}_{G}} = \int w_{IJ}(t,\bvect{x}) C_{I} C_{J} d^{N}x
\label{eqn:grandconstraint}
\end{equation}
with any positive definite matrix $w_{IJ}$.  Like $K_{mn}$,
the matrix $w_{IJ}$, can,
in principle, be a function of both space and time, and can have
off-diagonal entries.  In practice, however,
I have only used diagonal and constant
matrices.  Even in the diagonal case, the matrix is necessary because 
there is no a priori reason to believe that the constraints have the
same dimensions, and it also allows the different
constraints to be treated with different relative importance.
In addition, since there
is no natural scale for the grand constraint, one may always set one of the
coefficients $w_{IJ}$ in \eqnref{grandconstraint} to unity.
\end{subsection}

\end{section}

\begin{section}{Application to Maxwell}
\label{sec:maxwell}
This section takes Maxwell's equations as a concrete example of a
system of partial differential equations subject to a pointwise
constraint.  Following KWB \cite{baumgarte:em}, I consider two ways of
writing Maxwell's equations
for the vacuum
in terms of the vector potential $\vect{A}{i}$.
The first system, called System I, uses the evolution equations
\begin{eqnarray}
\pder{\vect{A}{i}}{t} & = & -\vect{E}{i} - \pder{\psi}{i} \label{eqn:simpA} \\
\pder{\vect{E}{i}}{t} & = & -\pdert{\vect{A}{i}}{j}{j} 
	+ \pdert{\vect{A}{j}}{j}{i} \label{eqn:simpE}
\end{eqnarray}
and the constraint 
\begin{equation}
C_{E} = \pder{\vect{E}{i}}{i} = 0. \label{eqn:simpC}
\end{equation}
The second system introduces the additional field $\Gamma$ 
defined by
\begin{equation}
\Gamma = \pder{\vect{A}{i}}{i} \label{eqn:Gammadef}
\end{equation}
to eliminate mixed derivatives in \eqnref{simpE}.  The evolution
equations for System II are 
\begin{eqnarray}
\pder{\vect{E}{i}}{t} & = & -\pdert{\vect{A}{i}}{j}{j} + \pder{\Gamma}{i}
	\label{eqn:bssnE} \\
\pder{\Gamma}{t} & = & -\pdert{\psi}{i}{i} \label{eqn:bssnGamma}
\end{eqnarray}
and \eqnref{simpA}.  Both systems use a gauge consistent with 
\begin{equation}
\pder{\psi}{t} = -\pder{\vect{A}{i}}{i} = -\Gamma \label{eqn:psievol}
\end{equation}
using the first equality for System I and the second for System II.

\begin{subsection}{System I Evolution Equations}
Having defined the systems, I would now like to calculate the terms required
for applying the constraint finding method to System I.  Here there is only
one constraint, \(C = C_{E} = \diverge{E}{i}\), which is zero-valued.  
It depends only
on the first derivatives of the electric field, therefore I need only to
calculate
\begin{equation}
\vder{\intsq{C^{2}}}{\vect{E}{i}(\bvect{x})}
	= -2 \pder{C_{E}}{i} \label{eqn:driveE}
\end{equation}
which modifies \eqnref{simpE}.  The new evolution equation for the 
electric field is
\begin{equation}
\pder{\vect{E}{i}}{t} = -\pdert{\vect{A}{i}}{k}{k} 
	+ \pder{\diverge{A}{k}}{i} + 2K_{E} \pder{C_{E}}{i}
	\label{eqn:fixedE}
\end{equation}
for an arbitrary positive $K_{E}$, while the other System I evolution
equations \eqnref{simpA} and \eqnref{psievol} remain unchanged.
\end{subsection}

\begin{subsection}{System II Evolution Equations}
System II, unlike System I, has two constraints that should be enforced,
the original constraint given by \eqnref{simpC} plus the definition of
$\Gamma$ in \eqnref{Gammadef}, rewritten as 
\begin{equation}
C_{\Gamma} = \diverge{A}{i} - \Gamma = 0
\label{eqn:GammaC}
\end{equation}
to make it zero-valued.  This provides more
freedom in constructing the grand constraint
\begin{equation}
C^{2} = C_{E}^{2} 
	+ w C_{\Gamma}^{2}
\end{equation}
where one does not necessarily have to treat the constraints on equal footing.
In this case, the total constraint depends additionally on $\Gamma$ and
first derivatives of the vector potential.  In addition to 
\eqnref{driveE}, which is still valid, I need
\begin{eqnarray}
\vder{\intsq{C^{2}}}{\vect{A}{i}(\bvect{x})} & = &
	-2w \pder{C_{\Gamma}}{i} \\
\vder{\intsq{C^{2}}}{\Gamma(\bvect{x})} & = &
	-2w C_{\Gamma}
\end{eqnarray}
to enforce the definition of $\Gamma$.

Applying these correction terms to the evolution equations gives the 
new equations of motion
\begin{eqnarray}
\pder{\vect{A}{i}}{t} & = & -\vect{E}{i} - \pder{\psi}{i}
	+ 2w K_{A} \pder{C_{\Gamma}}{i} \label{eqn:fixedA} \\
\pder{\vect{E}{i}}{t} & = & -\pdert{\vect{A}{i}}{k}{k} + \pder{\Gamma}{i}
	+ 2K_{E} \pder{C_{E}}{i} \label{eqn:fixedEII} \\
\pder{\Gamma}{t} & = & -\pdert{\psi}{k}{k} 
	+ 2w K_{\Gamma} C_{\Gamma} \label{eqn:fixedGamma}
\end{eqnarray}
which, combined with \eqnref{psievol}, form a complete system.  The constants 
$K_{A}$ and $K_{\Gamma}$ are arbitrary but positive.
\end{subsection}

\begin{subsection}{Propagation of Constraints}
In \cite{baumgarte:em}, KWB examined the evolution equation for $C_{E}$
in both systems.  They showed that for System I, the constraint does not
evolve in time, and that in System II, the constraint obeys a wave equation.
They did not have reason, however, to consider the time evolution of the 
secondary constraint $C_{\Gamma}$.  I extend their results here by showing that
the secondary constraint also satisfies a wave equation in the unmodified
case. Furthermore, I demonstrate the improved behavior of the 
constraints under the modifications proposed here.

Calculating the first time derivatives of the constraints is easily 
accomplished by taking the time derivatives of \eqnref{simpC} and
\eqnref{GammaC} and replacing the time derivatives that appear on the
right hand sides of the equations by the evolution equations 
\eqnref{fixedE}, \eqnref{fixedA}, and \eqnref{fixedGamma}.  This
gives the results
\begin{eqnarray}
\pder{C_{E}}{t} & = & -\laplace{\left[ C_{\Gamma} - 2K_{E} C_{E} \right]}{i}
	\label{eqn:evolCE} \\
\pder{C_{\Gamma}}{t} & = & -C_{E} 
	+ 2w \left[ K_{A} \laplace{C_{\Gamma}}{i} 
	- K_{\Gamma} C_{\Gamma} \right] \label{eqn:evolCGamma}
\end{eqnarray}
which can be viewed as valid for both systems if $C_{\Gamma}$ is taken
to be identically zero for System I.

To see that KWB have a wave equation for the secondary constraint, set
all of the $K$s to zero in (\ref{eqn:evolCE}--\ref{eqn:evolCGamma}) to
eliminate the modifications, and
take the time derivative of \eqnref{evolCGamma}.  The result
\begin{equation}
\laplace{C_{\Gamma}}{t} = \laplace{C_{\Gamma}}{i}
\end{equation}
follows immediately.

Of greater interest here, however, is an analysis on
the modified equations (\ref{eqn:evolCE}--\ref{eqn:evolCGamma}) in their
first derivative form.  The equations are linear, so they admit a
Fourier analysis by substituting a plane wave solution $e^{ikx}$ into the 
right hand sides.  After substituting, the resulting equations
\begin{eqnarray}
\pder{C_{E}}{t} & = & k^{2} C_{\Gamma} - 2K_{E}k^{2}C_{E} \\
\pder{C_{\Gamma}}{t} & = & 
	-C_{E} - 2w \left[ K_{A}k^{2} + K_{\Gamma} \right] C_{\Gamma}
\end{eqnarray}
retain the terms that gave the KWB wave equations for the constraints, but
have additional terms that look like they provide exponential decay.  This
system of equations is, in fact, simple enough for Maple to solve analytically
for general values of the $K$s in one dimension.  
The solution, which is too long to display in detail here, consists of a
sum of terms with the form
\begin{equation}
\exp \left[\left(-f^{+}_{1} 
	\pm \sqrt{\sigma \left[ \left(f^{-}_{1}\right)^{2} 
		- k^{2}\right]} \right) t \right] f_{2}
\label{eqn:waveform}
\end{equation}
where
\begin{equation}
f_{1}^{\pm}(k^{2})  =  (K_{E} \pm w K_{A}) k^{2} \pm w K_{\Gamma}
\end{equation}
\(\sigma = \pm 1\), and 
\(f_{2} = f_{2}(k, C_{i}(0,x))\)
is some simple function 
of $k$ and the initial values of the constraints.
Since \(f^{+}_{1}(k^{2}) > 0\) is manifestly positive and the
radical is either positive or pure imaginary, the only term of
this form that could cause anything other than exponential decay is
\(-f^{+}_{1} + \sqrt{\sigma [(f^{-}_{1})^{2} - k^{2}]}\) for
parameters where \(\sigma [(f^{-}_{1})^{2} - k^{2}] > 0\).  Simple
algebraic analysis, however, shows that even this term has an overall
minus sign, giving exponential decay.

In order to make the argument more concrete, I present here the 
$k=1$ solution  
\begin{eqnarray}
C_{E}(t,x) & = & e^{-3t} \left[ C_{E}(0,x) + S(x) t \right] \\
C_{\Gamma}(t,x) & = & e^{-3t} \left[ C_{\Gamma}(0,x) - S(x) t \right]
\end{eqnarray}
for which all of the $K$s are set equal to one.  Here 
\(S(x) = C_{E}(0,x) + C_{\Gamma}(0,x)\) is a short hand.  The general
solution can be requested from the author in the form of a 
Maple worksheet.
\end{subsection}

\begin{subsection}{Numerical Results}
My numerical experiments on these modified systems of equations
used an ICN integration scheme \cite{teuk:icn}, and a Courant factor of 1/2.  
The spatial domain ran from -6 to +6, with data stored on 99 points in each 
coordinate direction.
On the evolved fields (\td{E}{i}, \td{A}{i}, $\psi$, and $\Gamma$),
I imposed outgoing wave boundary conditions (see \cite{cactus:bound} for
implementation details), and on the constraints, I imposed $C_{I} = 0$
for applicable $I$.  
All runs were performed on a 500~MHz Digital Personal Workstation with 
1.5~GB of RAM.

For each system, I ran three parameter sets, one of which
reproduced the equations used by KWB.  The full definitions of all of the
parameter sets are found in \tabref{rundef}.
\begin{table}
\centering
\begin{tabular}{|l|c|c|c|c|} \hline
\textbf{Run} & $K_{E}$ & $K_{A}$ & $K_{\Gamma}$ & $w$ \\ \hline \hline
I-0  & 0 & - & - & - \\ \hline
I-1  & \sci{1}{-2} & - & - & - \\ \hline
I-2  & \sci{5}{-2} & - & - & - \\ \hline
II-0 & 0 & 0 & 0 & 1 \\ \hline
II-1 & \sci{5}{-3} & \sci{5}{-3} & \sci{5}{-3} & 1 \\ \hline
II-2 & \sci{1}{-2} & \sci{1}{-2} & \sci{1}{-2} & 1 \\ \hline
\end{tabular}
\caption{The parameters used for the various data runs done on the
two Maxwell systems are tabulated here.}
\label{tab:rundef}
\end{table}
Sensible values of the parameters were easily determined by trial and error.
Choosing the values too small, as expected, makes little difference in the
evolution, while choosing the values too large leads to numerical instabilities
during the transient period of the evolution \footnote{One possible 
explanation
for the instability associated with large parameters is that the
added terms modify the dispersion relationships for the various Fourier
modes, as seen in \eqnref{waveform}.  Large values of the parameters may
require adjustments to the Courant condition, which has not been analyzed 
here.}.

I followed KWB in using the analytic solution
\begin{eqnarray}
A^{\phi} & = & 0 \\
E^{\phi} & = & 8 \mathcal{A} \lambda^{2} r e^{-\lambda r^{2}} \sin \theta
\end{eqnarray}
of a toroidal dipole to generate the initial data.
The other components of the fields are
zero.  I chose \(\lambda = \mathcal{A} = 1\),
and the conversion from spherical to Cartesian coordinates was made in 
the code.

The results of the System I runs are summarized in \figref{sysI},
which shows a plot of $\normtwo{C_{E}}$ vs. $t$.
\begin{figure}
\centering
\epsfig{file=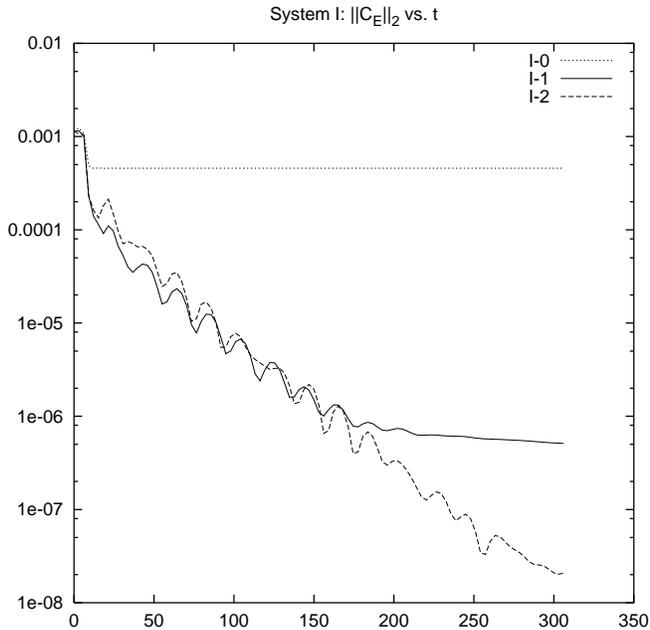, width=\columnwidth}
\caption{The $l_{2}$ norm of the primary constraint $C_{E}$ versus time $t$
for three test cases.  Case I-0 has no correction terms (\(K_{E} = 0 \)).
See \tabref{rundef} for the other parameter values.}
\label{fig:sysI}
\end{figure}
The I-0 (control) curve reproduces the findings of KWB that, after an 
initial transient, the $C_{E}$ constraint does not evolve in time.  The
I-1 and I-2 curves, representing different values (see \tabref{rundef})
of the parameter $K_{E}$, on the other hand, show a 
modulated exponential decay.  Eventually, around $t \approx 200$,
the I-1 case also stops decaying as rapidly, while the rapid exponential decay 
continues through the end of the run for I-2.

That the constraint in the modified case I-1 ceases to evolve at some point is 
consistent with 
\eqnref{gencorrection}, which implies that the constraints will cease to 
evolve when the first term balances with the second term.  From 
\eqnref{gencorrection}, one expects that this balance will be achieved
for smaller constraint violation when the driving parameter is larger, which
is consistent with the results shown in \figref{sysI}.

Looking at two dimensional slices of the constraint data at various times, also
suggests that the source of the modulation in the decay demonstrated
by I-1 and I-2 is fluctuations at
the boundary, possibly caused by the simple boundary condition applied there
on the constraints.  Significantly, these fluctuations are unable to penetrate
the interior of the computational domain, unlike many scenarios seen in
numerical relativity where noise from the boundary noticeably propagates
inward from the outer boundary, eventually killing the simulation.  Because
I am only interested in the Maxwell equations as a test-bed for the method,
and because the modified System I equations already perform orders of magnitude
better than their unmodified counter part, I have not pursued this point
further.

\figref{sysII} shows a 
plot of $\normtwo{C_{E}}$ vs. $t$ for the System II case.
\begin{figure}
\centering
\epsfig{file=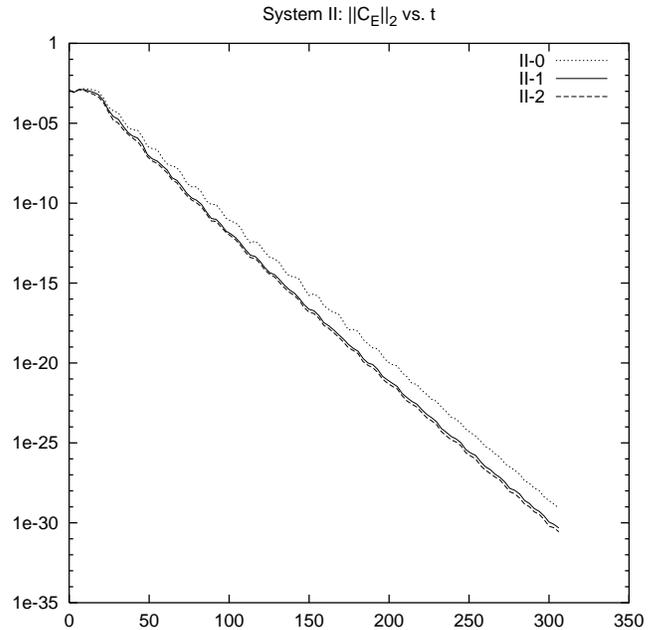, width=\columnwidth}
\caption{The $l_{2}$ norm of the primary constraint $C_{E}$ versus time $t$
for three test cases.  Case II-0 reproduces KWB.  See \tabref{rundef} for
the definitions of other parameters.}
\label{fig:sysII}
\end{figure}
Here again, the control run (II-0) reproduces the results of KWB, this time
showing exponential decay in the primary constraint.  Even with such an
ideal result in the unmodified case, the modified runs II-1 and II-2 show
improvement.  They represent runs with non-zero values (see \tabref{rundef})
of the various forcing parameters, and in these cases the constraint decays
exponentially, but with a smaller characteristic time.  

\figref{sysIIA} is a plot of $\normtwo{C_{\Gamma}}$ versus time $t$, and
shows similar behavior to the primary constraint in all three cases.
\begin{figure}
\centering
\epsfig{file=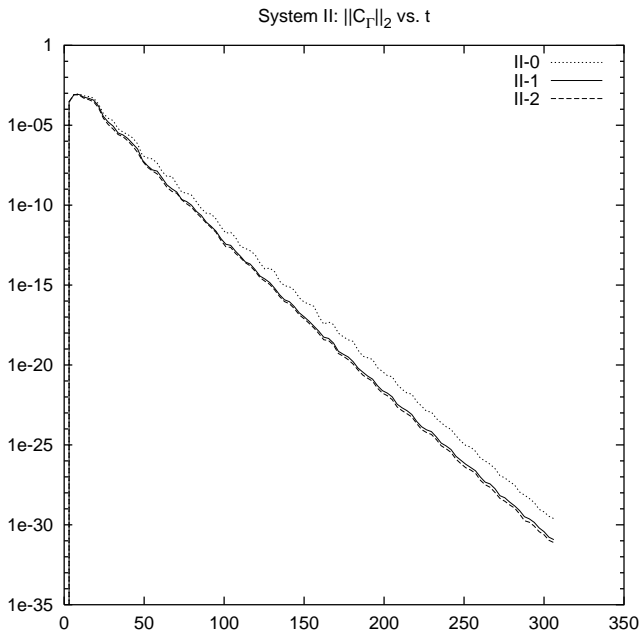, width=\columnwidth}
\caption{The $l_{2}$ norm of the secondary constraint $C_{\Gamma}$ 
versus time $t$ for three test cases.  Case II-0 is the result of KWB.
See \tabref{rundef} for other parameter values.  At $t=0$
the constraint is identically satisfied.}
\label{fig:sysIIA}
\end{figure}
The secondary constraint shows exponential decay in the unmodified II-0 case,
while showing a faster decay in the two modified runs, II-1 and II-2.  The
jump in the graph at $t=0$ occurs because the secondary constraint
is exactly satisfied in the initial data by construction.

It should be noted that one likely explanation for the extremely favorable
performance of the unmodified System II is that, since the constraints
satisfy a wave equation, constraint violations propagate off of the
grid.  This is supported by the data presented by KWB, showing that decay
rate of the constraint decreases when the outer boundary is moved farther
out \cite{baumgarte:em}.  The method presented here benefits from 
constraint violations propagating off of the grid
as well, but, in addition, attempts to damp the constraint violation
throughout the grid at all times.

In both systems, there is a practical limit to how large the forcing parameters
may be chosen without creating an instability.  I believe that its origin
can be seen in \eqnref{waveform}, which shows how the dispersion relation for
the Fourier components of the solutions is modified by the new terms.  This
modification is effectively a modification of the propagation speed of those
components, which, for a fixed Courant factor, can lead to numerical
instabilities in the numerical scheme.  When the forcing parameters are too
large, the code crashes.  The constraint blows up rapidly immediately before
such a crash, and such crashes usually occur during the transient period of
the numerical scheme.

\end{subsection}

\end{section}

\begin{section}{Conclusions and Discussion}
\label{sec:conclusions}
The method proposed here does not provide a formalism for any physical system.
It attempts to take any formalism and makes that formalism better.  
The analysis does not depend on the form of the equations and the 
arguments leading to the
modifications do not depend on the validity of any perturbative expansion.
The results in the case of the Maxwell equations are very encouraging
and consistent with the analytic predictions.

Should the results derived from Maxwell's equations generalize to the
Einstein equations, furthermore, this could prove extremely beneficial to the
long term stability of simulations in numerical relativity.  Fourier
analysis of the ADM system suggests that instabilities seen in various
3+1 formulations are triggered by exponentially growing modes not
seen in the initial data \cite{lsu:stab}.  On the other hand, at least for
the case of the Maxwell equations, the terms added via the method proposed here
lead to exponential damping terms in the evolution equations
for all Fourier modes.

It should be noted, however, that in some sense the Maxwell equations
were an optimal case.  Notice, by counting the number
of integrations by parts required to calculate the correction terms to the
evolution equations, that the highest order of spatial derivatives
in the new evolution equations will be the larger of (1) order of the original
equations and (2) twice the order of the constraints.  
This means that, while the order of the Maxwell evolution equations
was not changed, the Einstein equations
will change from second order to fourth order in space.  How this change of 
order effects the numeric solutions to the equations must be studied as this 
method is applied in the context of numerical relativity.
\end{section}

\begin{acknowledgments}
I would like to thank C. Misner, C. Schiff,
T. Jacobson, D. Mattingly, and V. Prieto-Gortcheva for 
comments on the ideas presented here and on the manuscript.
The code used in the numerical simulations made use of the Cactus computational
toolkit and a modified piece of code originally written by E. Schnetter for 
the Maya project.
This work was supported in part by NSF grant PHY-0071020.
\end{acknowledgments}

\bibliography{gr}

\end{document}